\documentclass[12pt,preprint]{aastex}

\def\lsim{\hbox{ \rlap{\raise 0.425ex\hbox{$<$}}\lower 0.65ex\hbox{$\sim$} }}
\def\gsim{\hbox{ \rlap{\raise 0.425ex\hbox{$>$}}\lower 0.65ex\hbox{$\sim$} }}

\def\opeqn{\begin{equation}}
\def\cleqn{\end{equation}}

\begin{document}

\title{ $n$-point Gravitational Lenses with $5(n-1)$ Images}
\author{Sun Hong Rhie }


\begin{abstract}
It has been conjectured (astro-ph/0103463) that a gravitational lens  
consisting of $n$ point masses can not produce more than $5(n-1)$ images 
as is known to be the case for $n = 2$ and $3$.
The reasoning is based on the number of finite limit points $2(n-1)$ which 
we believe to set the maximum number of positive images and the fact that
the number of negative images exceeds the number of positive images by $(n-1)$.
It has been known that an $n$-point lens system ($n\ge 3$) can produce $(3n+1)$ 
images and so has been an explicit lens configuration with $(3n+1)$ images.  
We start with the well-known $n$-point lens configuration that produces $(3n+1)$ images  
and produce $(2n-1)$ extra images by adding a small $(n+1)$-th mass so that  
the resulting $(n+1)$-point lens configuration has $(2n)$ discrete limit points
and produces $5n$ images of a source. 
It still remains to confirm in abstraction that the maximum number of positive image 
domains of a caustic domain is bounded by the number of the limit points.  
\end{abstract}


\clearpage

An $n$-point gravitational lens is an algebraic extension of the systems of masses 
whose gravitational effect is linear such that its lensing effect is given by the
following lens equation here written in terms of complex variables.   
\opeqn
 \omega = z - \sum_j^n {\epsilon_j\over \bar z_j} \equiv z - \bar F_n \qquad ; 
 \sum_j^n \epsilon_j = 1 \ ; \quad z_j \equiv z - x_j 
\label{eqLeq}
\cleqn 
where $\omega$ is the position of the light source and the total mass is normalized 
to 1 so that $\epsilon_j$ is the fractional mass of the point lens located at
$x_j$. The number of images is the number of solutions for $z$.  
Since the number of solutions of the equation changes where the Jacobian determinant
of the equation vanishes, the so-called critical curve where the Jacobian determinant
vanishes ($J(z, \bar z)=0$) consist of closed curves and the images are referred
to as positive or negative images depending on the sign of the $J$ value. 
Since the lens equation is ``almost analytic" (the derivatives are analytic),  
it is easier to handle the following $\kappa$ function. The critical condition is 
given by $|\kappa| = 1$. 
\opeqn
 \kappa \equiv {\partial\bar\omega\over\partial z} 
        = \sum_j^n {\epsilon_j\over z_j^2} \ ; \qquad J = 1 - |\kappa|^2 \le 1. 
\label{eqKappa}
\cleqn
It is worth emphasizing that the Jacobian determinant is bounded by 1. One immediate 
implication is that positive images ($J > 0$) are never demagnified ($|J|^{-1} \ge 1$).
Where $J=1$ is where the family of $J=$ constant contours contract to points, and
the limit points can be found by solving $\kappa=0$. 
Since $\kappa(z)=0:  z \not = \infty$ is an $2(n-1)$-order analytic equation, 
there are always $2(n-1)$ (finite) limit points but they can be degenerate.
If we let $z_j = r_j \exp i\theta_j$ and rewrite the equation $0 = \kappa $, 
\opeqn
  0 = \sum_j^n {\epsilon_j e^{i\theta_j}\over r_j^2} \ , 
\cleqn
it is clear that the limit points are where the Newtonian forces ($1/r^2$) 
due to the lens masses 
balance out and it corroborates why the images at the limit points are neither magnified 
nor demagnified ($|J|=1$). It also makes it easy to guess the rough positions
of the limit points in relation to the positions of the lenses. (The limit 
points can be considered 2-d Lagrange points.) For two masses, imagine
the line that connects them, and the limit points are on both sides of the line
closer to the smaller mass. For three masses, consider pairwise and their 
interactions, etc. 

It has been known in the cases of $n=2$ and $n=3$ that the discrete limit points  
can be the positive image positions of a source and in fact the source generates
the maximal number of images 5 and 10. It has been also found that the $(n+1)$ 
limit points of a set of $n$ equal masses equally spaced 
on a circle can be the positive image
positions of a source at the center of the circle when the radius of the circle
is properly chosen. If the radius of the circle is $a$ and one mass is on the
positive side of the real axis, the lens equation is given by $z = \bar F_n$ where
\opeqn
  F_n = {z^{n-1}\over z^n-a^n} \ ;  \qquad
  \kappa_n = {z^{n-2}(z^n+(n-1) a^n)\over (z^n-a^n)^2} \ .
\label{eqFn}
\cleqn
One limit point of degeneracy $(n-2)$ is at the center and is an image. The other
limit points are on a circle of radius $r_\ast = (n-1)^{1/n} a$ and are images 
when $r_\ast^2 = (n-1)/n$. Let's note that the limit points configuration is  
azimuthally shifted by $\pi/n$ in comparison to the configuration of the lens 
positions while both the angular distances between two neighboring lens masses 
and two neighboring limit points are $2\pi/n$.
\opeqn
 x_j = a e^{i{2k\pi\over n}} \ ; \quad
 z_\ast =  r_\ast e^{i{\pi\over n}} e^{i{2k\pi\over n}} \ : \qquad
 k = 0, 1, ... , n-1 
\cleqn
If we consider lines that go through the center, limit points and the lens positions 
define $n$ lines separated by $\pi/n$. If $n$ is odd, then on each line lie one
lens mass and two limit points one of which is at the center. On the real line,
in the order of increasing coordinate values, they may be represented as 
$\{ *  *  X \}$  where $*$ and $X$ denote limit points and lens positions respectively.
If $n$ is even, we need to consider two neighboring lines. For example, the real 
line contains two lens positions and one limit point: $\{ X  *  X \}$ and the 
neighboring line rotated by $\pi/n$ from the real line contains two limit points:
$\{ *  * \}$. 
The limit points are image positions if 
\opeqn
  r_\ast = \left({n-1\over n}\right)^{1\over 2} \ ; \quad
  a = (n-1)^{-{1\over n}} \left({n-1\over n} \right)^{1\over 2} \ .
\cleqn     
In order to find negative images, we need to solve the lens equation 
$z=\overline F_n$. 
\begin{enumerate}

\item
For odd $n ~(\ge 3)$, it suffices to consider the equation on 
the real line: $z = F_n$.  
There is always one solution at $z=0$ and that is the positive image at the 
limit point. The other solutions satisfy the following. 
\opeqn
  0 =  h  \equiv z^n - z^{n-2} - a^n 
\label{eqH}
\cleqn     
This equation has three solutions: one on the positive real axis and two on the
negative real axis. That is because $h(z_+) h(z_-) < 0$ where $z_\pm$ are the
two solutions to$0=h^\prime$.
\opeqn
  z_\pm = \pm \left({n-2\over n}\right)^{1\over 2}  \ ; \quad
  h(z_+) h(z_-) = - {(n-2)^{n-2}\over n^n} 
      \left(4 - \left({n-1\over n-2}\right)^{n-2}\right)  
\cleqn
The factor in the last parenthesis monotonically decreases with $n$ but remains
positive. It is $2$ for $n=2$ and converges to $(3-e) \approx 0.28$ as 
$n \rightarrow \infty$.  One of the solutions on the negative real axis is the 
limit point $z = - r_\ast$ and so a positive image. The sequence of images and
lens positions in the increasing order of the coordinate values is 
$\{ *  \circ  *  X  \circ \}$ where $\circ$ denotes the positions of negative images.
In the case of $n=3$, the negative image solutions are at 
$(1 \pm\sqrt{3})/\sqrt{6}$. 
Their $J$ values are $(-2)$ and the well-known sum rule works out: 
$\sum_{images} J^{-1} =1$. The inverse Jacobian of the two negative images and
the limit point at $z = -r_\ast$ cancel out, and the total is given by the value
of the limit point at the center. That is not the case for $n > 3$. 

\item
For even $n ~(\ge 4)$, the real line has only two (negative) images 
because $h(z_+)=h(z_-)$ and so $ h(z_+) h(z_-) > 0$.      
The sequence of images and lens positions is given by $\{ \circ X * X \circ \}$.
On a neighboring line, we set $z = t \exp(i\pi/n): t \not= 0 $ and find two 
solutions: $\{ * \circ  *  \circ * \}$.  
\end{enumerate}

Now we add a small mass $n\epsilon$ at the center. 
The thought is that dropping in a lensing mass that defines the position of a pole
($J = -\infty$) at the current limit point that defines a zero ($\kappa=0$ or $J=1$)
should alter the "topographic structure" of the image plane defined by the function 
$J$ drastically enough to change the number of images but should be perturbatively 
traceable by keeping the mass small.
\opeqn
 F_{n+} = {z^n - \epsilon^\prime a^n \over z(z^n-a^n)} \ ;  \qquad
  \kappa_{n+} = {z^{2n}+((n-1)-(n+1)\epsilon^\prime)z^n+\epsilon^\prime a^{2n}
                \over z^2(z^n-a^n)^2} \ 
\label{eqFn+}
\cleqn
If $z$ is an image of the ring lens configuration, $z = \overline{F_n(z)}$, then 
we look for solutions $z+\eta$ that are images of the new configuration with an
extra mass in the lowest but sufficient order in $\eta$: 
$z+\eta = \overline{F_{n+}(z+\eta)}$.   
\begin{enumerate}
\item
The limit point equation near the center shows $n$ limit points near the center. 
\opeqn
 0 = \epsilon^\prime a^{2n} + a^n\eta^n (n-1) \quad \Rightarrow \quad   
  \eta = \left(\epsilon^\prime\over n^2 \right)^{1\over n} 
          e^{i{\pi\over n}} e^{i{2k\pi\over n}}: \ \ k=0, 1, ... , (n-1)
\label{eqNewlimit}
\cleqn
The radius $|\eta|$ depends on one $n$-th power of the small mass $\epsilon^\prime$ 
and one can say that the original limit point is split into $n$ new limit points. 
The total number of limit points increases by $(n-1)$ to $(2n)$ as expected for 
non-degenerate limit points of a $(n+1)$-point lens. 
\item
 In order to investigate the positions of the images, let's consider the real
line first. Near the center, the lens equation becomes 
$0 = -\eta^2 +\epsilon^\prime$ and two new images are formed.
\opeqn
 \eta = \pm\sqrt{\epsilon^\prime} \ ; \quad
  J(\eta) = - \eta^n (n-1)^2 \left({n\over n-1}\right)^{n\over 2}  
\cleqn
If $n$ is odd, the image on the positive real axis (in the direction of the lens
position with respect to the center) is a negative image and the image on the 
negative real axis is a positive image. The new positive image is in fact 
facilitated by the new limit point on the negative real axis found  
in equation (\ref{eqNewlimit}). We should note that the new positive image is
not at the position of the new limit point but in the ``positive domain"
newly defined by the new limit point. 
If $n$ is even, the both are negative images, which is to be expected  
because the lens masses (where $J =-\infty$) are along the real line.   
\item
When $n$ is even, lens axes (lines through lens positions and the center) and 
``limit point axes" (lines through limit points and the center) are separate
as we discussed. The set of ``limit point axes" is rotated by $\pi/n$ from the
set of lens axis. It is simple to find that there are two positive images
along each ``limit point axis".
\opeqn
  \eta = \pm\sqrt{\epsilon^\prime} e^{i{\pi\over n}} \ ; \quad  
  J(\eta) = -\eta^n (n-1)^2 \left({n\over n-1}\right)^{n\over 2}
\cleqn
\item
The images and limit points away from the center shift their positions slightly
but their counts and parities remain the same.
The positive images depart from the limit point positions. Off-center limit 
points move toward the center with increasing $\epsilon^\prime$. Images move
away from the center. Thus, the effect of the small $(n+1)$-th mass 
$\epsilon^\prime$ added to the center is to increase the number of the limit
points from $(n+1)$ to $(2n)$ and to increase the number of images by $(2n-1)$
from $(3n+1)$ to $(5n)$.     
\end{enumerate}

In conclusion, we have shown via explicit constructions that the maximum 
number of images of the $n$-point gravitational lens is no less than
$5(n-1)$. It has been suspected that $5(n-1)$ is in fact the maximum based
on the maximum number of discrete limit points $2(n-1)$. We mentioned
above an evident correlation between new positive images and new limit points.  
If the number of positive images is bounded by the number of (finite) limit 
points ($n\ge2$) as we conjectured, then the total number of images can not 
exceed $5(n-1)$ because of the relation between the numbers of positive and
negative images. It still remains to find an abstract analysis to confirm that 
the maximum number of positive image domains of a caustic domain is bounded 
by the maximum number of the limit points. Of course, refutation has not been
ruled out, but we recommend not to bet on it.   

\end{document}